\documentstyle[12pt,epsfig]{article}
\begin{document}
\title{Neutrinos and the Standard Model}
\author{Barry R. Holstein\\
Department of Physics-LGRT\\
University of Massachusetts\\
Amherst, MA  01003}
\maketitle
\begin{abstract}
Since their "discovery" by Pauli in 1930, neutrinos have played 
a key part in confirmation of the structure of the
standard model of strong and electroweak interactions.  After reviewing 
ways in which this has been manifested in the past, we discuss
areas in which neutrinos continue to play this role.
\end{abstract}
\newpage

\section{Introduction}
The neutrino is a particle whose impact on contemporary physics far outweighs 
its (possible) negligible mass.  Indeed even the layman has long been 
fascinated by a particle which is (essentially) massless, chargeless, and 
which can pass through the earth without interaction---{\it cf.} 
the poem by
John Updike written nearly four decades ago (allegedly when he became bored 
during a Harvard physics lecture)\cite{upd}.
\begin{center}
Neutrinos they are very small\\
$\,\,\,\,\,$They have no charge and have no mass\\
And do not interact at all.....\\
\end{center}
At this meeting, we will be hearing from many experts on contemporary aspects
of neutrinos, especially having to do with their role in astrophysics and
cosmology.  I shall not attempt to compete with these experts, but rather will
discuss ways in which the neutrino has 
impacted and continues to affect our understanding of the structure of the
standard model.  After a brief historical introduction, I will emphasize ways
in which the neutrino has affected the evolution of standard model structure 
even from the beginning and then will mention areas of contemporary
physics wherein the neutrino continues to play a key role. 

\section{Neutrino History}

The "discovery" of the neutrino is quite different from that of its sibling
leptons in that its existence was inferred nearly three decades 
before its actual experimental confirmation.  Indeed it was Pauli 
who in 1930 postulated 
the existence of a light neutral particle inside the 
nucleus---"not larger than 0.01 proton 
masses"---and called by him the "neutron," in order to explain why nuclear 
beta decay was observed to have a continuous (three-body) rather
than discrete (two-body) electron energy spectrum\cite{pau}.  This 
issue of the spectrum was so troublesome at the time
that no less an authority than Niels Bohr had speculated that it might
be necessary to abandon the idea of energy conservation, except in a 
statistical sense, when considering subatomic processes such as beta 
decay\cite{boh}.   Of course, there was a serious problem with Pauli's
suggestion, in that a quick uncertainty principle estimate shows that
a particle this light has a position uncertainty $\Delta x\sim 1/m_\nu\sim
400$ A and could not therefore be confined within the nuclear volume.  
This problem was
solved by Fermi, who renamed this particle the "neutrino" and proposed 
his famous field theory of
beta decay
\begin{equation}
{\cal H}_w={G_F\over \sqrt{2}}\psi_p^\dagger{\cal O}_\mu\psi_n
\psi_e^\dagger{\cal O}^\mu\psi_\nu+h.c.
\end{equation}
wherein this particle does not exist inside the nucleus but rather is
created as a byproduct of the decay itself\cite{fer}.  The one unknown
constant $G_F$ can be determined from the neutron lifetime via
\begin{eqnarray}
\Gamma_n&=&\left({G_F\over \sqrt{2}}\right)^2\int{d^3p_e\over (2\pi)^3}
{d^3p_\nu\over (2\pi)^3}2\pi\delta(M_n-M_p-E_e-E_\nu)|{\cal M}_w|^2\nonumber\\
&=&{G_F^2\over 4\pi^3}\int_{m_e}^{M_n-M_p}dE_eE_ep_e(M_n-M_p-E_e)^2|{\cal M}_w|^2
\nonumber\\
&\simeq&4.59\times 10^{-19}{\rm GeV}^5G_F^2|{\cal M}_w|^2=887\pm 2\, {\rm sec}
\end{eqnarray}
which yields $G_F\simeq 10^{-5}M_p^2$.  This was all very convincing and
Bohr soon became a believer, acknowledging 
"Finally, it may be remarked that the 
grounds for serious doubts as regards the strict validity of the conservation
laws in the problem of the emission of $\beta$-rays from atomic nuclei are
now largely removed by the suggestive agreement between the rapidly 
increasing experimental evidence regarding $\beta$-ray phenomena and the
consequences of the neutrino hypothesis of Pauli so remarkably developed in
Fermi's theory"\cite{boh1}.

This is all well and good but it is one thing to postulate the existence of
the neutrino and quite another thing to actually detect it.  The problem
lies in the size of the weak coupling inferred from beta decay.  One can
easily calculate a resulting neutrino scattering cross section as
\begin{eqnarray}
\sigma_\nu&=&\left({G_F\over \sqrt{2}}\right)^2\int{d^3p_e\over (2\pi)^3}
2\pi\delta(M_p+E_{\bar{\nu}}-M_n-E_e)|{\cal M}_w|^2\nonumber\\
&\sim&{G_F^2\over 2\pi}p_eE_e|{\cal M}_w|^2\sim 10^{-44}{\rm cm}^2\,
\,\,{\rm at}\,\,\,E_{\bar{\nu}}=1\,{\rm MeV}
\end{eqnarray} 
The mean free path passing through a medium of earthlike
density is then expected to be
\begin{equation}
\Delta x\sim 1/(\rho\sigma_\nu)\sim 10^{21}\,\,{\rm cm}
\end{equation}
or $10^{10}$ earth radii!!  The solution to this problem, of course, is to
get lots of neutrinos, many target nuclei, or better yet {\it both}!
One of the original ideas conceived by Cowan and Reines to this problem of
getting many neutrinos on target was to set off a nuclear bomb near an
underground detector\cite{lanl}.  They soon had a more realistic thought, 
however, and decided
to place the detector near a reactor.  After original work at the Hanford
site, they moved their base of operations to Savannah River and in 1956
were able to announce the unambiguous discovery of the neutrino via the
reaction $\bar{\nu}_e+p\rightarrow n+e^+$\cite{rei}.

In retrospect, this discovery took place in the middle of a tremendous
amount of seminal work, which led within a decade to the picture which we
now call the standard model of weak and electromagnetic interactions.  This
included
\begin{itemize}

\item [i)] Suggestion of parity violation by Lee and Yang\cite{lee} and its
subsequent experimental confirmation\cite{wua};

\item [ii)] Postulation of the V-A structure of the weak current by Feynman
and Gell-Mann\cite{fey} and its confirmation;

\item [iii)] Development of the quark model by Gell-Mann and Zweig\cite{gmz};

\item [iv)] Proposal of quark mixing by Cabibbo\cite{cab};

\item [v)] Discovery of the standard electroweak model by Weinberg and 
Salam\cite{wei}.

\end{itemize}

By 1967 then we already had what has become one of the most successful 
theories in modern physics.  In this picture the neutrino plays a pivotal
role and has at least three fundamental aspects which have been subjected to
extensive experimental tests:

\begin{itemize}

\item [i)] Chirality:  Because of the $1+\gamma_5$ structure of the weak
interaction, the neutrino (antineutrino) must be purely left-handed 
(right-handed).

\item [ii)] Dirac Character:  The neutrino is predicted of Dirac character,
possessing a distinct antiparticle, rather then a Majorana particle which is
its own antiparticle.

\item [iii)] Mass:  The neutrino is massless, implying that there is 
no lepton analog to the CKM mixing occuring in the charged weak current.

\end{itemize}

Each of these predictions has been studied over the years and we
now have a sizable data base of experimental evidence involving each issue.
I will summarize each in turn:

\subsection{Chirality}

The prediction of definite chirality was first studied by Goldhaber, Grodzins,
and Sunyar in 1958 via electron capture on ${}^{152}Eu$ to an excited
state of ${}^{152}Sm$ followed by its subsequent radiative decay to the ground
state\cite{ggs}.  By studying those photons 
which are emitted opposite to the direction
of the outgoing neutrinos one can show that the photon and neutrino
helicities must be identical.  Thus one can study the neutrino helicity by
measuring that of the photon.  When this was done the authors announced 
that "our result seems compatible with ... 100\% negative helicity of 
neutrinos emitted in orbital $e^-$ capture," although they did not really
quantify this assertion.

Since that time the chirality issue has been extensively studied.  The
way one does this is to postulate a form for the charged current
weak interaction which includes right-handed components.  A typical form
for the semileptonic interaction is\cite{hot} 
\begin{equation}
{\cal L}={G_F\cos\theta\over \sqrt{2}}\left[(V_\mu-\rho A_\mu)(v^\mu-a^\mu)
+(xV_\mu+y\rho A_\mu)(v^\mu+a^\mu)\right]
\end{equation}
where $V_\mu,A_\mu$ ($v^\mu,a^\mu$) are hadronic (leptonic) weak currents
respectively.  
Here $\rho=(1-x)/(1-y)$ with $x,y$ being parameters which characterize the
possible existence of right-handed effects.  In a minimal left-right model of
spontaneous symmetry breaking, we would identify
\begin{equation}
x\simeq \delta-\zeta,\quad y\simeq \delta+\zeta
\end{equation}
where $\delta=M_1^2/M_2^2$ measures the ratio of (predominantly) 
left- and right-handed
gauge boson masses and $\zeta$ is the mixing angle defined via
$W_1=W_L\cos\zeta-W_R\sin\zeta$.  The tightest limits on 
$x,y$ come from precise beta decay studies.  Examples include measuring
the longitudinal polarization of the electron, which is given by
$P_L\simeq \beta(1-2y^2)$ or of the asymmetry parameter in the decay of 
polarized nuclei, which for neutron decay has the form 
\begin{equation}
A=2{g_A(g_A+g_V)-yg_A(yg_A+xg_V)\over g_V^2+3g_A^2+(x^2g_V^2+3y^2g_A^2)}
\end{equation}
Over the years a series of careful studies has produced the limits shown in
Figure 1, which generally limit $x,y$ at the several percent level\cite{deu}.
(The reason that generally tenth of a per cent precision in beta decay 
measurements results in only several percent limits on $x,y$ 
is due to the feature
that left and right handed currents do not interfere, so that any 
deviations from standard model predictions are quadratic in $x,y$ as can
be seen above.)

\begin{figure}
\centerline{\epsfig{file=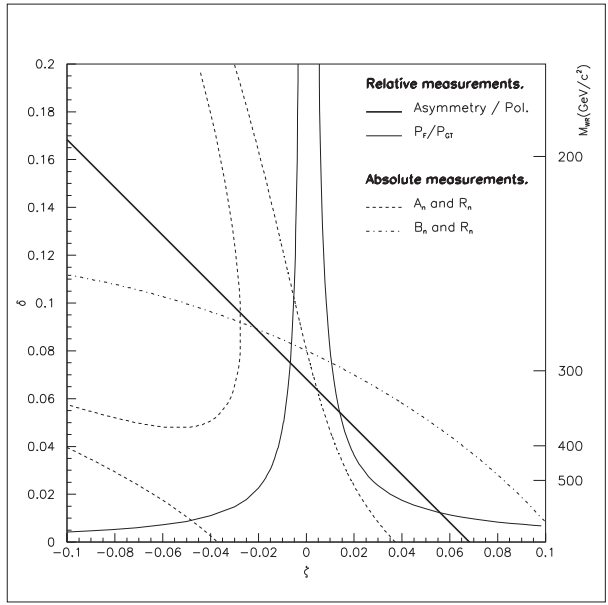,height=8cm}}
\caption{Present experimental limits on right-handed parameters from
beta decay experiments.}
\end{figure}

\subsection{Dirac Character}

The Dirac character of the neutrino has also been extensively examined, and
Frank Avignone has, of course, been extensively involved in such studies.  
Naively one might think that the issue would already be clear from the 
feature that while the reaction 
\begin{equation}
\nu_e+p\rightarrow n+e^+\nonumber
\end{equation}
does not occur while
\begin{equation}
\nu_e+p\rightarrow p+e^-\nonumber\\
\end{equation}
does.  Equivalently the absence of neutrinoless double beta decay
\begin{equation}
(A,Z)\rightarrow (A,Z+2)+e^-+e^-\nonumber\\
\end{equation}
which can occur via sequential beta decay accompanied by the exchange of
a virtual neutrino (=antineutrino) would seem to argue strongly against a
Majorana character.  However, both of these arguments are
blunted if the neutrino has definite helicity, as experimentally seems
to be the case.  Indeed then even if the neutrino has a Majorana character,
it has the wrong helicity to bring about the scattering or double beta
decay reactions above, so that their experimental absence does not bear
on the Dirac vs. Majorana issue.  On the other hand, if the neutrino is
Majorana and has a small mass, so that helicity is not definite, then
neutrinoless double beta decay is possible and it is experiment involving
${}^{76}Ge$ which Frank pioneered and has been doing for many years.  
Just this week a new limit on the Majorana mass of $<m_\mu^M>\leq 0.2$ eV has
been announced from such measurements\cite{cern}.

\subsection{Neutrino Mass}

The issue of whether the neutrino has a mass is clearly a fundamental one.  In 
the standard model the absence of mass is due to Ockham's Razor---{\it i.e.}, 
the standard
model uses only the {\it minimal} number of components.  Since a right handed
neutrino structure is not necessary, the standard model assumes its absence 
and, as it requires both a left and right handed component in
order to generate a mass, the neutrino is predicted to be massless.  This
prediction is one which has been under experimental scrutiny for many years.
(Even Fermi in his original paper suggested examination of this
issue by looking at the electron energy dependence of beta decay spectra
near the endpoint\cite{fer}.)
Generally most such studies have
utilized ${}^3H$ due to its low---18.6 KeV---endpoint, since that maximizes
the interesting component of the electron spectrum, and such
studies have become increasingly precise.  An early value by Hamilton, Alford,
and Gross placed the upper limit at 250 KeV\cite{hag}, 
which in 1972 was lowered to   
60 eV by Bergkvist\cite{ber}.  In 1980 Lubimov announced a nonzero value
$14 eV\leq m_{\bar{\nu}}\leq$ 46 eV, which set off a firestorm of new 
work\cite{lub}.
Present experiments do not not find evidence for a nonzero 
neutrino mass and upper bounds
have been place at 9.3 eV by Robertson et al.\cite{rob} and at 
7.2 eV by Weinheimer et al.\cite{wein}, so that the 
Lubimov value has been superceded.  Work continues on
such direct mass measurements.  Some interesting new ideas have
been discussed at this workshop, including use of Rhenium, with an 
endpoint energy even lower than that of ${}^3H$ and the use of bolometric
methods to detect the electron.

In is interesting to note in this regard, that in the middle of this
intense activity to measure neutrino mass, an event occurred which bears 
on this issue and which allows a simple limit to be set which is nearly 
comparable to those obtained from these careful spectral studies---SN1987a,
which blazed into the sky on February 27, 1987 and was observed not only
optically, but also by neutrino detectors in the US and Japan.  If the 
neutrinos emitted by the supernova were massive, then the 
velocity would be $v\simeq 1-m_\nu^2/(2E_\nu^2)$ and 
the most energetic neutrinos would reach the earth first.  It is easy to
estimate the time difference as
\begin{equation}
{\delta t\over t}\sim{\delta v\over v}\sim{m_\nu^2\over E_\nu^2}{\delta E_\nu
\over E_\nu}
\end{equation}
and the time gap between the arrival
of the high and low energy neutrinos could then be used to measure this mass.
Experimentally, the $\sim$ 10 MeV neutrinos arrived over a $\sim$ 10 second
time interval after travelling a distance of 165,000 light years 
from the supernova in the Large Magellanic Cloud 
but a time-energy correlation was 
not observed.   One can then easily set a limit on the mass---
\begin{equation}
m_\nu\leq E_\nu\left({\delta t\over t}{E_\nu\over \delta E_\nu}
\right)^{1\over 2}\sim 10\,{\rm sec.}\left(10\,s\over 10^{13}\,s
\right)^{1\over 2}\sim 10\,{\rm eV}
\end{equation}
A more careful analysis sets the upper bound at about 20 eV.  It is astounding
to me that the relatively trivial analysis given above from an event occurring
long before the dawn of civilization is able to set a limit on neutrino mass 
comparable to that obtained from years of precise experimental studies!

Of course, I have summarized here only the {\it direct} mass 
measurements.   Simultaneously, a series of experiments involving a search
for neutrino mixing has been underway.  Such mixing is prohibited in the
absence of mass since neutrino identities could just be reassigned.  The 
recent announcements of mixing signals from solar, accelerator, and atmospheric
measurements then clearly, if comfirmed, indicates the existence of 
neutrino mass.   Since
this will be the subject of many talks during this workshop, I will not
here summarize this data but instead will move on to discuss aspects of
neutrino physics which are not as well known, but which have a bearing on
contemporary physics issues.

\section{Contemporary Issues}

Above we have seen how the neutrino has played an essential role in
development of the structure of the standard model.  In this section, I 
argue that this is still going on and discuss ways in which
neutrino interactions are involved in a number of the central issues in
contemprary physics.  In this discussion, I will not emphasize some of the
more traditional ways in which this is manifested---{\it e.g.} 
\begin{itemize}

\item [i)] use of neutrino scattering in order to study the $Q^2$ evolution
of deep inelastic structure functions as a test of perturbative QCD,

\item [ii)] use of such deep inelastic structure functions in order to
check the validity of various sum rules, such as the Adler sum rule
\begin{equation}
1=\int_0^1 {dx\over 2x}(F_2^{\nu n}(x,q^2)-F_2^{\nu p}(x,q^2)),
\end{equation} 
\end{itemize}
since these are fairly well known.  Instead I will discuss three 
lesser known applications wherein neutrino studies have a bearing on
interesting standard model issues.

\subsection{Goldberger-Treiman Discrepancy}

One of the important features of QCD is its (broken) chiral symmetry, from
which follows the existence of the Goldberger-Treiman (GT) relation, which
connects the strong pion-nucleon coupling $g_{\pi NN}$ and the axial
coupling $g_A(0)$ measured in neutron beta decay\cite{gtr}, 
\begin{equation}
M_Ng_A(0)=F_\pi g_{\pi NN}(0)\label{eq:gt}
\end{equation}
where $F_\pi=92.3$ MeV is the pion decay constant.
One subtlety associated with Eq. \ref{eq:gt} is that the pi-nucleon coupling
is evaluated {\it not} at the physical point---$q^2=m_\pi^2$---but 
rather at the
unphysical value---$q^2=0$.  In fact when the physical coupling is used,
one expects a {\it violation} of the GT identity and this is often
showcased by quoting the so-called Goldberger-Treiman discrepancy 
\begin{equation}
\Delta_\pi=1-{g_A(0)M_N\over g_{\pi NN}(m_\pi^2)F_\pi}
\end{equation}
Strictly speaking the value of $\Delta_\pi$ is given by a chiral 
counterterm, but in a reasonable model one would expect $g_\pi NN(q^2)$ to 
vary with $q^2$ in essentially the same way as its weak analog 
$g_A(q^2)$.  In this way one finds
\begin{equation}
\Delta_\pi=1-{g_A(0)\over g_A(m_\pi^2)}={1\over 6}r_A^2m_\pi^2\simeq 0.034
\label{eq:ns}
\end{equation}
where $r_A=0.65\pm 0.03$ fm is the axial charge radius 
measured in charged current neutrino
scattering\cite{neu}.\footnote{Here the axial charge radius is defined
via
\begin{equation}
g_A(q^2)=g_A(0)(1+{r_A^2\over 6}q^2+\ldots).
\end{equation}}

An alternative approach is to utilize the Dashen-Weinstein
relation 
\begin{equation}
\Delta_\pi={\sqrt{3}m_\pi^2F_K\over 2m_K^2F_\pi}\left({g_{\Lambda KN}\over
g_{\pi NN}}\Delta_K^\Lambda-{1\over \sqrt{6}}{g_{\Sigma KN}\over g_{\pi NN}}
\Delta_K^\Sigma\right)
\end{equation}
which predicts the pionic GT discrepancy in terms of its kaonic analogs 
involving $\Lambda$ and $\Sigma$ couplings respectively\cite{dw}.  The 
original proof of this result argued that it was valid up to terms second 
order in chiral symmetry breaking.  However, recently
it has been shown by Goity et al. that in heavy baryon chiral 
perturbation theory any such difference can arise only at ${\cal O}(p^5)$
or higher\cite{goi}.  Although the strong $\Lambda,\,\Sigma$ couplings are
not well determined, the predictions are only weakly dependent upon thse
quantities.  Thus
one finds the relatively robust prediction
\begin{equation}
\Delta_\pi({\rm Dashen-Weinstein})=0.017
\end{equation}
in good agreement with that expected from neutrino scattering results.

Now what does experiment say?  The problem here is that while the pion
decay constant, the nucleon mass, and the weak axial coupling are all
well known, there is still considerable debate about the value of the
size of the pion nucleon coupling constant.  A recent analysis of 
$NN,N\bar{N},\pi N$ data by the Nijmegen group yields the value
$g_{\pi NN}(m_\pi^2)=13.05\pm 0.08$\cite{nij} and a VPI analysis yields
similar results\cite{vpi}.  However, a significantly larger 
number---$g_{\pi NN}(m_\pi^2)=13.65\pm 0.30$---has been found by Bugg and
Macleidt\cite{bm} and by Loiseau\cite{loi}.  When these values are 
used in order to calculate the 
GT discrepancy, we find
\begin{eqnarray}
\Delta_\pi=0.014\pm 0.006\quad {\rm if} \quad g_{\pi NN}=13.05\pm 0.08
\nonumber\\
\Delta_\pi=0.056\pm 0.020\quad {\rm if} \quad g_{\pi NN}=13.65\pm 0.30
\end{eqnarray}
The neutrino scattering number Eq. \ref{eq:ns} then comes right in the 
middle, while the Dashen-Weinstein analysis strongly supports the lower value
of $g_{\pi NN}$.  

\subsection{Axial Charge Radius}

A second interesting application of neutrino scattering results is 
associated with confirmation of a prediction of chiral perturbation theory
and therefore of QCD.  In order to understand this point, we return to
the early days of current algebra and PCAC and a low energy theorem derived
by Nambu and Schrauner, which argues that the axial charge 
radius
may be obtained via measurement of the isospin odd $E_{0+}$ multipole in
threshold electroproduction via\cite{nam}
\begin{equation}
E_{0+}^{(-)}(m_\pi=0,k^2)={eg_A\over 8\pi F_\pi}\left(1+{k^2\over 6}r_A^2
+{k^2\over 4M_N^2}(\kappa_V+{1\over 2})+{\cal O}(k^3)\right)
\end{equation} 
In this way one has determined the value $r_A=0.59\pm 0.05$ fm,\cite{ep} 
differing from the number $r_A=0.65\pm 0.03$ fm found via direct neutrino 
scattering measurements.  Although the discrepancy is only at the one sigma
level, it is interesting that recent calculations by Bernard, Kaiser, and
Meissner in heavy baryon chiral perturbation theory have shown that the 
old low energy theorem is incorrect and that there exists an additional
contribution coming from so-called triangle diagrams, which predicts a
difference between the axial charge radius as measured in neutrino scattering
and that from electroproduction\cite{bkm}
\begin{equation}
r_A^2({\rm elec.})=r_A^2({\rm neu})-{3\over 64F_\pi^2}({12\over \pi^2}-1)
\end{equation} 
The 0.046 fm$^2$ difference predicted from chiral symmetry agrees well in
size and sign with that seen experimentally.

\subsection{Nucleon Strangeness Content}

My final example has to do with the subject of strangeness content of the
nucleon, which is one of intense current interest.  One of the early studies
of such matters is the paper of Donoghue and Nappi\cite{dn}.  The idea 
here is that one expects that in the limit of vanishing quark masses 
the nucleon mass should approach some nonzero value $M_0$.  
In the real world, with
nonzero quark mass, the nucleon mass is modified to become
\begin{equation}
M_N=M_0+\sigma_s+\sigma
\end{equation}  
where, defining $\hat{m}=(m_u+m_d)/2$, 
\begin{equation}
\sigma_s={1\over 2M_N}<N|m_s\bar{s}s|N>,\quad \sigma={1\over 2M_N}
<N|\hat{m}(\bar{u}u+\bar{d}d)|N>
\end{equation}
are the contributions to the nucleon mass from strange, non-strange quarks
respectively.  One constraint in this regard comes from study of the hyperon
masses, which yields
\begin{eqnarray}
\delta&=&{\hat{m}\over 2M_N}<N|\bar{u}u+\bar{d}d-2\bar{s}s|N>\nonumber\\
&=&{3\over 2}{m_\pi^2\over m_K^2-m_\pi^2}(M_\Xi-M_\Lambda)\simeq 25 MeV
\end{eqnarray}
and increases to about 35 MeV when higher order chiral 
corrections are included.
A second constraint comes from analysis of $\pi N$ scattering, which 
says that $\sigma$ can be extracted directly if an isospin even combination
of amplitudes could be extrapolated via dispersion relations 
to the (unphysical) Cheng-Dashen point
\begin{equation}
F_\pi^2D^{(+)}(s=M_N^2,t=m_\pi^2)=\sigma
\end{equation}
When this is done the result comes out to
be $\sim$60 MeV, which is lowered to about 45 MeV by higher order chiral
corrections.  If $<N|\bar{s}s|N>=0$, as might be expected from a naive
valence quark picture, then we would expect the value coming from the
hyperon mass limit and that extracted from $\pi N$ scattering to agree.  The
fact that they do not can be explained by postulating the existence of a 
moderate strange quark matrix element
\begin{equation}
f={<N|\bar{s}s|N>\over <N|\bar{u}u+\bar{d}d+\bar{s}s|N>}\simeq 0.1
\end{equation}
implying $M_0\simeq 765$ Mev and $\sigma_s\simeq 130$ MeV, which
seem quite reasonable.  However, recent analyses have suggested a 
rather larger value of the sigma term, leading to $f\simeq 0.2$, $M_0\simeq
500$ MeV and $\sigma_s\simeq$ 375 MeV, which appear somewhat too large.  This
problem is ongoing.

In any case it is of interest to study the possibility of a significant
strange quark matrix element in other contexts.  One quantity which has
been extensively studied is the nucleon electromagnetic matrix element, which
has the form
\begin{eqnarray}
&&<N|V_\mu^{em}|N>=<N|{2\over 3}\bar{u}\gamma_\mu u-
{1\over 3}\bar{d}\gamma_\mu d
-{1\over 3}\bar{s}\gamma_\mu s|N>\nonumber\\
&=&\bar{u}(p')\left[\gamma_\mu(F_1^{ns}(q^2)+F_1^s(q^2))
-{i\over 2M_N}\sigma_{\mu\nu}q^\nu(F_2^{ns}(q^2)+F_2^s(q^2))\right]
u(p)
\end{eqnarray} 
and one can look for the existence of strange quark pieces
$F_1^s(q^2),F_2^s(q^2)$ in parity-violating electron scattering.  This
has been done in the forward direction by the HAPPEX experiment at
JLab\cite{hap} and in the backward direction by the SAMPLE experiment at 
MIT-Bates\cite{sam}.
The HAPPEX result is consistent with $r_1^s=0$, while the Bates result
seems to indicate a small positive value for $\mu_s$.  

Another probe comes from the realm of deep inelastic electron
scattering wherein,
defining the quark helicity content $\Delta q$ via
\begin{equation}
\Delta q\sigma_\mu=<p,\sigma|\bar{q}\gamma_\mu\gamma_5q|p,\sigma>
\end{equation}
one has the constraint 
\begin{equation}
\int_0^1 dxg_1(x)={1\over 2}\left[{4\over 9}\Delta u+{1\over 9}\Delta d
+{1\over 9}\Delta s\right](1-{\alpha_s(q^2)\over \pi})
\end{equation}
When combined with the Bjorken sum rule and its SU(3) generalization
\begin{eqnarray}
\Delta u-\Delta d&=&g_A(0)=F+D\nonumber\\
\Delta u+\Delta d-2\Delta s&=&3F-D
\end{eqnarray}
one finds the solution $\Delta u=0.81,\,\Delta d=-0.42,\,
\Delta s=-0.11$, indicating a small negative value for the strange
matrix element.  

So far, these results have nothing to do with our
main focus, which is neutrinos.  However, we note that there exists an 
alternative probe for such a strange
matrix element which is accessed via neutral current neutrino scattering.
The point here is that the form of the standard model axial current is
\begin{equation}
<N|A_\mu^Z|N>={1\over 2}<N|\bar{u}\gamma_\mu\gamma_5u-\bar{d}
\gamma_\mu\gamma_5d-\bar{s}\gamma_\mu\gamma_5s|N>
\end{equation}
which is purely isovector in the case that the strange matrix element 
vanishes and can be therefore be 
exactly predicted from the known charged current axial matrix element.
This experiment was performed at BNL and yielded a result\cite{bnl}
\begin{equation}
\Delta s= -0.15\pm 0.09
\end{equation}
consistent with that found from the deep inelastic sector, but a more
precise value is needed.

\section{Conclusion}

We have argued above that the neutrino has played and continues
to play an important role in the development of the standard model.
In the past such studies contributed to the now accepted picture of 
the weak interaction.  Present work looks for small deviations
from this structure.  However, we have also seen how neutrino experiments
bear on a number of issues of great interest in contemporary physics and
I suspect that neutrino measurements will continue to be exciting far
into the new millenium.

\begin{center}
{\bf Acknowledgement}
\end{center}

It is a pleasure to acknowledge the hospitality of University of South
Carolina and the organizers of this meeting.  This work was supported in
part by the National Science Foundation.

\end{document}